\documentclass[aps, prd, twocolumn,nofootinbib,groupedaddress,floatfix]{revtex4-2}
\usepackage{amsmath, amssymb, amsthm, float,mathtools}
\usepackage[urlcolor=blue,colorlinks,breaklinks=true]{hyperref}

\PassOptionsToPackage{obeyspaces,spaces,hyphens}{url}

\usepackage{graphicx}
\usepackage{dcolumn}
\usepackage{bm}
\usepackage{xcolor} 

\usepackage{todonotes}

\newcommand{\be}[0]{\begin{equation}}
\newcommand{\bq}[0]{\begin{eqnarray}}
\newcommand{\ee}[0]{\end{equation}}
\newcommand{\eq}[0]{\end{eqnarray}}
\newcommand{\cc}[0]{{\tilde c}}
\newcommand{\vv}[0]{\bar{v}}
\newcommand{\lp}[0]{\bar{L}}
\newcommand{\xd}[0]{\dot{\mathbf{x}}}
\newcommand{\xx}[0]{{\mathbf{x}}}
\newcommand{\mc}[1]{\mathcal{#1}}
\newcommand{\Cb}[0]{\hat{C}}
\newcommand{\kv}[0]{k(\vv)}

\begin{document}

\title{A velocity-dependent two-scale model for cosmic string networks with small-scale structure}\

\author{T. O. Miranda}
 \email[Electronic address: ]{teresa.miranda99@gmail.com}
\affiliation{Departamento de F\'{\i}sica e Astronomia, Faculdade de Ci\^encias, Universidade do Porto, Rua do Campo Alegre 687, PT4169-007 Porto, Portugal}
\affiliation{Instituto de Astrof\'{\i}sica e Ci\^encias do Espa{\c c}o, Universidade do Porto, CAUP, Rua das Estrelas, PT4150-762 Porto, Portugal}
\author{L. Sousa}%
 \email[Electronic address: ]{lara.sousa@astro.up.pt}
 \affiliation{Instituto de Astrof\'{\i}sica e Ci\^encias do Espa{\c c}o, Universidade do Porto, CAUP, Rua das Estrelas, PT4150-762 Porto, Portugal}
\affiliation{Departamento de F\'{\i}sica e Astronomia, Faculdade de Ci\^encias, Universidade do Porto, Rua do Campo Alegre 687, PT4169-007 Porto, Portugal}

\date{\today}

\begin{abstract}
 We develop a semi-analytical model to describe the cosmological evolution of networks of cosmic strings with small-scale structure, by extending the velocity-dependent one-scale model to include an additional lengthscale describing the typical interkink density. We study the impact of the different physical processes involved in the production and removal of small-scale structure from cosmic strings on the attainment of a full linear scaling regime, in which the characteristic lengths of the network and of small-scale structure evolve proportionally to physical time and the root-mean-squared velocity of the network remains constant. We find, using this novel velocity-dependent two-scale model, that quite generally small-scale structure does not prevent the attainment of a linear scaling regime since, even if not enough kinks are carried away when loops are chopped from the network, gravitational backreaction is generally enough to ensure that the interkink density scales. We find, however, that this regime is characterized by a smaller energy density and root-mean-squared velocity when compared to strings without small-scale structure and that this reduction may be significant when scaling is maintained by gravitational backreaction. In this case, we also find that, before reaching full scaling, the network should evolve in a transient quasi-scaling regime, in which its evolution is very similar to that of cosmic strings without small-scale structure.
\end{abstract}

\maketitle

\section{Introduction}
Cosmic strings provide us with a valuable observational window into high-energy physics. These line-like topological defects are predicted to form in symmetry-breaking phase transitions in a large variety of early universe scenarios~\cite{kibble_1976,Witten:1984eb,Davis:1986xc,Vilenkin:2000jqa,Sarangi:2002yt,Jeannerot:2003qv,Dror:2019syi}, but they are, in general, expected to survive until the present time. As a result, cosmic strings may potentially leave characteristic imprints on various observational probes --- including the Cosmic Microwave Background~\cite{Sazhina:2019ebp,Raidal:2026cpb,Caloni:2026dyu}, the primordial Stochastic Gravitational Wave Background~\cite{blanco2014number,Sousa:2013aaa,Cui:2017ufi} and large-scale structure surveys~\cite{Hernandez:2021vqh,Jiao:2023wcn} --- that enable us to probe them observationally and to uncover the physics of the early universe through them. A detailed understanding of the cosmological evolution of cosmic string networks, however, is crucial to perform accurate characterizations of their observational signatures and to use this observational data to its full potential.

Cosmic string networks have been extensively studied using both numerical simulations~\cite{Lorenz:2010sm,blanco2011large,Hindmarsh:2017qff} and analytical models~\cite{Kibble:1984hp,kibble_q,three_scale_model,martins_VOS,martins_k(v)_chopping,Correia:2021tok} and a general picture about their cosmological evolution has emerged: string networks evolve towards a linear scaling regime in which their energy density remains a fixed fraction of the background energy density. This means that a single lengthscale --- the characteristic length $L=(\rho/\mu)^{1/2}$, where $\rho$ is the cosmic string energy density and $\mu$ is cosmic string mass per unit length --- may be sufficient to describe their cosmological evolution on sufficiently large scales. Numerical simulations, however, also revealed that cosmic strings accumulate, as a result of string intersections and intercommutations, structure on scales much smaller than this characteristic length. This small-scale structure may have a significant impact not only on the evolution of the network, as it may provide an additional energy loss channel for long strings~\cite{Quashnock:1990qy,Sakellariadou:1990ne,Hindmarsh:1990xi} and have an important role in loop production~\cite{Polchinski:2007rg,Dubath:2007mf}, but also on their observational signatures~\cite{Copeland:1999gn,Pogosian:1999np,Auclair:2020oww,Silva:2023diq}. In particular, it is not yet clear whether small-scale structure will scale or build up on cosmic strings and potentially prevent the network from fully reaching a linear scaling regime. 

Numerical simulations of cosmic string networks, due to resolution and dynamical range limitations and because they generally do not include gravitational backreaction, are yet to provide a definite answer to these open questions. As an alternative, several semi-analytical models~\cite{kibble_q,small_scale_Structure,allen_caldwell_q,three_scale_model,Martins:2014kda}, often including multiple lengthscales, have been developed in the literature to study and quantify the impact of small-scale structure on string networks, but these models are usually complex and include a large number of phenomenological parameters. Here, we build up on this work and develop a semi-analytical model that describes the evolution of the cosmic string networks using two distinct lengthscales, the characteristic length $L$ and a characteristic length for small-scale structure $l_k$, but that also treats the root-mean-squared velocity of the network as a dynamical variable. This results in a model that is simpler than many others models in the literature and that has a significantly smaller number of free parameters, thus making the study of the evolution of cosmic string networks with small-scale structure analytically a more tractable problem. Despite this, this velocity-dependent two scales model for cosmic string networks with small-scale structure produces results that are similar to those obtained with more complex models that exist in the literature. In particular, it indicates that, although small-scale structure should not prevent the network from attaining a full linear scaling regime, it may have a significant impact on their cosmological evolution and lead to a decrease of their energy density at the present time.

This paper is organized as follows. In Sec.~\ref{sec:VOS} we revisit the Velocity-dependent One-Scale (VOS) model for the evolution of cosmic string networks without small-scale structure. In Sec.~\ref{sec:sss}, we derive an evolution equation for the linear kink density of long strings and discuss the impact of the different physical processes involved in the attainment of a linear scaling regime. In Sec.~\ref{sec:v2s}, we extend the VOS model to allow for the description of small-scale structure, by incorporating the evolution equation for the characteristic lengthscale of small-scale structure derived in the previous section. We also study the attainment of a full linear scaling regime in this model and derive analytical expressions to characterize this regime. We then discuss the results in Sec.~\ref{sec:disc}.

Throughout this paper, we will use natural units with $c=\hbar=1$, where $c$ is the speed of light in vacuum and $\hbar$ is the reduced Planck constant. Moreover, unless stated otherwise, we assume a $\Lambda$CDM cosmological background, with the cosmological parameters determined by Planck 2018 data~\cite{Planck:2018vyg}: the values of the density parameters for radiation, matter and dark energy at the present time are respectively given by $\Omega_{\rm r} = 9.1476\cdot 10^{-5}$, $\Omega_{\rm m} = 0.308$, $\Omega_\Lambda=1-\Omega_{\rm r} - \Omega_{\rm m}$ and the Hubble constant is $H_0=2.13 \cdot h \cdot 10^{-33} \, \rm eV$, with $h=0.678$.

\section{The velocity-dependent One-scale model}\label{sec:VOS}
In a Friedmann-Lemaître-Robertson-Walker (FLRW) universe, described by the line element
\begin{equation}
    ds^2=dt^2-a^2(t)d\mathbf{x}\cdot d\mathbf{x}\,,
\end{equation}
(where $a$ is the cosmological scale factor, $t$ is physical time and $\mathbf{x}$ are comoving coordinates), a cosmic string network may be assumed to be homogeneous and isotropic on sufficiently large scales. The Velocity-dependent One-Scale (VOS) Model describes macroscopically the dynamics of such a cosmic string network, thus enabling for a quantitative description of its cosmological evolution. It makes use of two dynamic variables: the root-mean-squared (RMS) velocity, $\bar{v}$, and the characteristic length, $L \equiv (\mu/\rho)^{1/2}$, where $\mu$ is the cosmic string tension and $\rho$ the average energy density of long strings. In this context, this lengthscale also describes approximately the mean distance between two neighboring strings and their typical curvature radius. Assuming that the cosmic strings are infinitely thin, evolution equations for these macroscopic variables may be obtained by averaging the Nambu-Goto equations of motion \cite{martins_VOS, martins_k(v)_chopping} over the whole network. There are, however, two additional important mechanisms to consider. When strings collide, they exchange partners and reconnect, in a process usually known as intercommutation. When two strings simply intercommute, there is no immediate energy loss by the network. However, when a closed loop of string is formed, it detaches from the network, evolving separately, and decays, therefore resulting in an energy loss. The rate of energy loss into loops is given by \cite{kibble_1976}:
\begin{equation}\label{energy loss to loops}
    \left(\frac{d\rho}{dt}\right)_{\rm loops}  = \tilde{c}\bar{v}\frac{\rho}{L} \ ,
\end{equation}
where $\tilde{c}$ is a phenomenological parameter that characterizes the loop-chopping efficiency. Numerical simulations show that $\tilde{c} = 0.23 \pm 0.04$ is a good fit in both radiation and matter dominated eras \cite{martins_k(v)_chopping}. The second aspect one needs to consider is that, during their evolution, cosmic strings will scatter off the relativistic particles of the background plasma, resulting in a frictional force that effectively decelerates the strings in the early stages of their evolution and leads to considerable additional energy loss during this stage. The impact of friction can be described by adding a characteristic friction lengthscale, $l_f \propto a^3$~\cite{Vilenkin:2000jqa,martins_VOS}. However, since frictional forces become negligible as the Universe expands and becomes less dense, they only affect the dynamics of the network very early on. For the remainder of this paper, we will then consider only the evolution of the networks in the frictionless epoch. The VOS Model equations are then~\cite{martins_VOS,martins_k(v)_chopping}: 
\begin{align}\label{VOS - L}
    \frac{dL}{dt} &= (1+\bar{v}^2)HL + \frac{1}{2}\tilde{c}\bar{v} \\ \label{VOS-v}
\frac{d\bar{v}}{dt}& = (1-\bar{v}^2)\left[\frac{k(\bar{v})}{L}-2H\vv\right] \ ,
\end{align}
where $H=\dot{a}/a$ is the Hubble parameter and
\begin{equation}\label{k(v)}
    k(\bar{v}) = \frac{2\sqrt{2}}{\pi} (1-\bar{v}^2) (1+2\sqrt{2} \bar{v}^3) \frac{1-8\bar{v}^6}{1+8\bar{v}^6} 
\end{equation}
is a dimensionless curvature/momentum parameter~\cite{martins_k(v)_chopping}. It was first introduced to provide a phenomenological description of the effect of small-scale structure on the dynamics of the cosmic string network, but, as we shall see, introducing other lengthscales may be necessary to provide a more precise description of its effects. We discuss this parameter further in Sec.~\ref{sec:disc}.

The VOS Model predicts the existence of a linear scaling regime, in which the energy density of the network remains constant relative to the background energy density, characterized by a linear attractor solution of the form: 
\begin{equation} \label{linear scaling regime}
    \frac{L}{t} = \xi =  \sqrt{\frac{k(\bar{v})(k(\bar{v})+\tilde{c})}{4\beta(1-\beta)}}   \quad \text{and } \quad\bar{v} = \sqrt{\frac{k(\bar{v})(1-\beta)}{(k(\bar{v})+\tilde{c})\beta}}  \ ,
\end{equation}
where we are assuming that the evolution of the Universe is ruled by a power law of the form of $a \propto t^{\beta}$, with constant $0<\beta<1$. This linear scaling regime only occurs for a constant value of $\beta$, meaning that it is only attainable either deep in the radiation-dominated or the matter-dominated eras, breaking off during the radiation-matter transition and once the cosmological constant starts to become relevant to the dynamics of the Universe. Note however that, since the matter era does not last long enough for the cosmic string network to re-establish scaling, in general, it should not be expected to be in a linear scaling regime after the onset of the matter era.

\section{Evolution of Small-Scale Structure on Cosmic Strings}\label{sec:sss}
The production and evolution of small-scale structure on cosmic string networks has been studied and modeled by different groups throughout the years~\cite{Quashnock:1990qy,Carter:1990nb,allen_caldwell_q,small_scale_Structure,three_scale_model, kibble_q, Carter:1994zs,Martins:2014kda,Vieira:2016vht,Almeida:2021ihc,Almeida:2022qbl}. Although these studies sometimes treat the physical processes involved differently and use different methodologies, a general picture for the evolution of small-scale structure has emerged. Discontinuities in the string tangent, known as kinks, are created when intercommutations occur, independently of loops being produced or not. These kinks, however, decay by emitting Gravitational Waves (GWs) and due to the stretching caused by the expansion of the Universe and  may also be carried away by loops, as they detach from the network.

In this section, we will study the impact of each of these processes on the evolution of the characteristic lengthscale of small-scale structure, $l_k$, describing the average distance between kinks. The objective is to understand their impact and relevance for the achievement of a linear scaling regime for $l_k$, in order to develop an effective model to describe the evolution of small-scale structure that combines the relevant features of the different models proposed in the literature. Following~\cite{small_scale_Structure, allen_caldwell_q}, we will often express our results in terms of the linear kink density, $K=1/l_k$, describing the number of kinks per unit physical length of string. Bearing in mind that the number density of kinks is given by $K/L^2$, we should have that~\cite{allen_caldwell_q}

\be
\frac{d}{dt}\left(\frac{K}{L^2}\right)=\left[\frac{dn_c}{dt}-\frac{dn_r}{dt}\right]-3H\frac{K}{L^2}\,,
\label{eq:K-master}
\ee
where $n_c$ and $n_r$ are, respectively, the number density of kinks created on and removed from the strings. Moreover, the last term in this expression accounts for the dilution of the kink number density caused by the expansion of the background.

\subsection{Impact of Loop Production}\label{subsec:loopprod}
Loop chopping plays a crucial role in the evolution of small-scale structure and, as a result, characterizing the rate of loop production accurately is essential for the development of our model. Let $n_{\rm loops}$ be the total number density of loops produced until a time  $t$. The loop production rate is given by:
\begin{equation}\label{dnloopdt}
    \frac{dn_{\rm loops}}{dt} = \frac{1}{E_{\rm loop}}\left(\frac{d\rho}{dt}\right)_{\rm loops} \ ,
\end{equation}
where $d\rho_{\rm loops}/dt$ is given by Eq. (\ref{energy loss to loops}) and $E_{\rm loop}=\mu \ell$ is the energy of loops at the moment of formation, which is determined by their physical length at that instant of time $\ell$.  We will assume, from this point forward, that loops are all created with the same length\footnote{Note that, although realistically one does not expect all loops to be created with exactly the same length, the results obtained by making this assumption can be used to get results in situations where the length of loops at birth follows any distribution \cite{Blanco_Pillado_2014,analytical_approx}.}, $\ell$, as is often assumed in computations of the stochastic gravitational wave background generated by cosmic string loops. In these computations, it is also commonly assumed that $\ell$ is determined by physical time (see e.g. ~\cite{Sanidas:2012ee,Blanco_Pillado_2014,Cui:2017ufi}), since the string network is often assumed to always evolve in a linear scaling regime. However, this was shown~\cite{Sousa:2013aaa} to lead to a significant underestimation of the number of loops created after the radiation-matter transition, wherein networks can no longer be assumed to be in a linear scaling regime. Since we want our study to apply to cosmic string networks in a realistic cosmological background, we will assume that the length of loops is a fixed fraction of the characteristic lengthscale as proposed in~\cite{Sousa:2013aaa}: $\ell = \alpha L$, where $\alpha$ is a constant such that $0\le\alpha\le1$.  This assumption is supported by numerical simulations of Nambu-Goto string networks~\cite{Blanco_Pillado_2014}, which show that the large loops produced by the network are well described by $\alpha \sim 0.34$ both in the radiation and matter eras. Later, in Sec.~\ref{sec:modloop}, we will relax this assumption and consider the possibility that $\ell$ is determined by the characteristic lengthscale of small scale structure, $l_k$, as well. 

In \cite{Quashnock:1990qy,small_scale_Structure}, the authors propose that the rate of kink creation is proportional to the rate of loop production:
\begin{equation} \label{kink creation}
    \frac{dn_c}{dt} = F \frac{dn_{\rm loops}}{dt} \ ,
\end{equation}
where $F$ is the average number of kinks created on the long string network for each loop formed. Although two kinks are created on each long string when strings intercommute, multiple reconnections may be necessary before a loop is produced. We should then have that $F \ge 2$~\cite{small_scale_Structure}.

Every time a loop detaches from the network, a length $\ell$ of string is removed, alongside $N_r = K\ell$ kinks. Therefore, every time a loop detaches, $N_r = \alpha L/l_k$ kinks are removed from the network. However, in \cite{kibble_q}, the authors proposed that loop production could be more likely in regions that have a larger density of kinks --- since strings are more curved in these regions and move with higher speeds --- and, as a result, loops may carry away more kinks than this naive expectation may indicate. This effect may be modeled by introducing a parameter, $q$, that characterizes the \textit{relative kinkiness} between the loop and the long string, so that the linear kink density of the loops is larger than that of the long strings by a factor of $q$. This means that, when a loop detaches from the network, it will remove $qK\ell$ kinks, instead of $K\ell$ and then, the rate at which kinks are removed from the network is then~\cite{small_scale_Structure,Quashnock:1990qy,kibble_q}:
\begin{equation}\label{dn_r/dt}
    \frac{dn_r}{dt} = \ell q K(t) \frac{dn_{\rm loops}}{dt} \,.
\end{equation}

Using Eq.~\eqref{eq:K-master}, we may write
\begin{equation}\label{diff equation lkd with q}
\begin{split}
    \frac{dK}{dt}&= [F-\alpha q L K]L^2\frac{dn_{\rm loops}}{dt} +\frac{2K}{L}\frac{dL}{dt}-3HK =\\
    &= F\frac{\tilde{c}\bar{v}}{\alpha L^2}- q\cc\vv\frac{K}{L} +\frac{2K}{L}\frac{dL}{dt}-3HK.
\end{split}
\end{equation}

Assuming that the long string network is in a linear scaling regime (i.e. that $\xi$ and $\bar{v}$ remain constant), this equation admits an analytical solution of the form:

\begin{equation}\label{case 1 analytical solution}
    K(t) = \frac{\mc{A}}{\mathcal{N} t} \left[\left(\frac{t}{t_i}\right)^{\mathcal{N}} - 1\right] \ ,
\end{equation}
where we defined $\mc{A}=F\cc\vv/(\alpha\xi^2)$ and $\mathcal{N}=3-(\cc \vv q)/\xi-3\beta$ and we have assumed that the network is formed at a time $t_i$ without small-scale structure (i.e., $K(t_i)=0$). The characteristic lengthscale of small-scale structure is in linear scaling regime if $l_k \propto t$ (or, equivalently, $K \propto 1/t$), which only occurs if $\mathcal{N}<0$. We may then see that, if $q$ is larger than a critical value
\be 
q_c^*= \frac{3\xi}{\tilde{c}\bar{v}}(1-\beta)\,,
\label{eq:qc1}
\ee
the removal of kinks from the network by loops is enough to ensure linear scaling, as was shown in \cite{small_scale_Structure}. Note that, using Eq.~\eqref{linear scaling regime}, we may rewrite this as

\be 
q_c^*=1+\frac{\xi}{\cc \vv}\left[(1-\beta)+2\beta\vv^2\right]\,,
\ee 
which shows that $q_c^*>1$ for all values of $\beta$. This demonstrates that the attainment of a linear scaling regime by the characteristic length of small-scale structure is only possible if loops are indeed typically kinkier than the long strings. As a matter of fact, they would have to be significantly kinkier: in the radiation era, we have $q_{c,r}^* \sim 2.67$, while $q_{c,m}^* \sim 4.67$ in the matter era. When $q<q_c^*$, the linear kink density will evolve as $K \propto t^{\mathcal{N}-1}$ and, therefore, the amount of small-scale structure on cosmic strings grows continuously over time. Also, for $q=q_c^*$, it is also not possible to attain a linear scaling regime for $l_k$ and we have that $t K \propto \log(t/t_i)$.

\subsection{Impact of Non-Loop Forming Intercommutations}\label{subsec:inter}
Kinks are formed in the network every time strings interact, even when that interaction does not lead to the creation of a loop. A question then arises: do the intercommutations that do not form loops have a significant impact on the creation of small-scale structure? The effect of these non-loop-forming intercommutations was treated in two different ways in the literature. In~\cite{Quashnock:1990qy,small_scale_Structure}, the authors assume that the total number of intercommutations is also proportional to the number of loops produced and therefore this effect may be described by considering a value of $F$ larger than 2. Refs. ~\cite{kibble_q, three_scale_model}, on the other hand, describe non-loop-forming intercommutations by including an additional term in Eq.~(\ref{kink creation}): 
\begin{equation}\label{intercom}
   \frac{dn_c}{dt} = F\frac{dn_{\rm loops}}{dt} + \frac{dn_{int}}{dt}\equiv F\frac{dn_{\rm loops}}{dt}+\frac{\chi(\bar{v})}{L^4}\,,
\end{equation}
where $n_{int}$ is the number density of kinks added to the network by non-loop forming intercommutations, and $\chi(\bar{v}) = 2\bar{v}(1 - \bar{v}^2)/\pi$ describes the probability of long string intercommutation.

Let us start by considering the latter case and, for consistency, let us set $F=2$ to avoid considering the same effect twice. Eq. (\ref{diff equation lkd with q}) now becomes:
\begin{equation}\label{diff K with inter}
    \frac{dK}{dt}= [2-\alpha q L K]L^2\frac{dn_{\rm loops}}{dt} +\frac{2K}{L}\frac{dL}{dt}-3HK + \frac{\chi(\bar{v})}{L^2} \ .
\end{equation}
If one now assumes a linear scaling regime, it is straightforward to see that Eq. (\ref{diff K with inter}) also admits a solution of the form of Eq.~\eqref{case 1 analytical solution}, but now 
\be 
F \longrightarrow F(\vv)\equiv 2 + \frac{\chi(\bar{v})\alpha}{\tilde{c}\vv}=2\left[1 + \frac{\alpha(1-\bar{v}^2)}{\pi\tilde{c}}\right]\,.
\label{eq:Fv}
\ee 
which in a linear scaling regime is also a constant. This clearly shows that, in a linear scaling regime, the effect of non-loop-forming intercommutations is indeed similar to taking a larger value of $F$: the critical value of $q$ remains unchanged, but there is an increase of the linear kink density as $K \propto F$. However, because the scaling value of $\vv$ is typically dependent on the expansion rate $\beta$ (\textit{cf}. Eq.~\eqref{linear scaling regime}), no single value of $F$ can be chosen throughout the evolution (especially when we consider, as in Sec.~\ref{sec:v2s}, a realistic cosmological background and treat $\vv$ as a dynamical variable). Note also that $F(\vv)$ is also dependent on $\alpha$. As may be seen from Eq.~(\ref{eq:Fv}), if the loops produced are small (and then many loop-forming intercommutations occur), the impact of these additional intercommutations is negligible and we have $F\approx 2$. For larger values of $\alpha$, however, this is no longer the case and the linear kink density may be indeed be enhanced. For these reasons, from this point forward, we will then consider by effect by replacing $F$ by $F(\vv)$.

\subsection{Impact of Expansion}\label{sec:expansion}
Let us now consider the impact of the stretching of kinks caused by the Hubble expansion of the background. This effect may be described by a term of the form\cite{kibble_q}:
\begin{equation}
    \left.\frac{1}{K} \frac{dK}{dt}\right|_{\rm stretching}= -2H(1-2\bar{v}^2)\,,
\end{equation}
so that Eq. (\ref{diff equation lkd with q}) then becomes
\bq
\frac{dK}{dt} & = & F(\vv)\frac{\tilde{c}\bar{v}}{\alpha L^2}- q\cc\vv\frac{K}{L} + \frac{2K}{L}\frac{dL}{dt} - \label{k eq with all decay}\\ 
& - & HK\left[3+2(1-2\bar{v}^2)\right] \nonumber\ .
\eq
Assuming that the long string network is in a linear scaling regime, once again we find that the solution to this equation is given by~\eqref{case 1 analytical solution}, but now we have 

\be
\label{eq:N2}
\mathcal{N}=3-q\frac{\cc \vv}{\xi}-\beta\left[3-2(1-2\vv^2)\right]\,.
\ee
The linear kink density may then attain a linear scaling regime if the relative kinkiness exceeds a new critical value given by

\be
q_c \equiv q_c^* - \frac{2\beta\xi}{\tilde{c}\bar{v}}(1-2\vv^2)=1+\frac{\xi}{\cc\vv}\left[1-3\beta(1-2\vv^2)\right]\,.
\label{eq:critical2}
\ee
The decay of kinks caused by stretching then facilitates the attainment of a linear scaling regime, as it leads to a reduction in the critical value of $q$. Its impact becomes more and more relevant as the expansion rate increases and, in fact, the values of the critical kinkiness in the radiation and matter era become similar: $q_{c,r}=2.45$ and $q_{c,m}=2.67$. Interestingly, for fast enough expansion rates with $\beta > 0.73$, a linear scaling regime may now be attained even for $q=1$.

\subsection{Impact of Kink Decay due to the emission of gravitational waves}\label{subsec:kink-decay}
The decrease in the linear kink density caused by the emission of gravitational radiation is of the form~\cite{allen_caldwell_q, Quashnock:1990qy}:
\begin{equation}
    \left.\frac{1}{K}\frac{dK}{dt}\right|_{\rm gw}= - \hat{C} \Gamma G\mu K = - \hat{\Gamma}G\mu K \ ,
\end{equation}
where $\hat{\Gamma} \equiv \Gamma\hat{C}$, $\Gamma \simeq 50$~\cite{Quashnock:1990wv,Casper:1995ub,Blanco-Pillado:2015ana} is the power emitted in GWs by long strings in units of $G\mu^2$ and $\hat{C} > 1$ was introduced to account for the impact of gravitational backreaction, which should lead to a smoothing of the kinks as they emit GWs and may make their decay more efficient \cite{three_scale_model}. The larger the value of $\hat{C}$ considered, the faster kinks will be smoothed out, and, therefore, $\hat{\Gamma}$ could be interpreted as an effective value for the GW emission efficiency, accounting for the effects of gravitational backreaction. 

Considering now both the decay by Hubble stretching and by GW emission, Eq. (\ref{dn_r/dt}) becomes
\begin{equation}
    \frac{dn_r}{dt} = \ell K\frac{dn_{loop}}{dt} + \frac{dn_{GW}}{dt} + \frac{dn_H}{dt} \ ,
\end{equation}
and so, Eq. (\ref{diff equation lkd with q}) may be written as 

\bq
\label{k eq with all decay}
\frac{dK}{dt} & = & F(\vv)\frac{\tilde{c}\bar{v}}{\alpha L^2}- q\cc\vv\frac{K}{L} + \frac{2K}{L}\frac{dL}{dt} -\label{k eq with all decay}\\ 
& - & HK\left[3+2(1-2\bar{v}^2)\right] -\hat{C}\Gamma G\mu K^2\nonumber\ .
\eq

Let us start by looking for the conditions that need to be satisfied for the linear kink density to reach a linear scaling regime, by solving this equation analytically. In this case, again assuming that $K(t_i)=0$ and that the long string network remains in a linear scaling regime, the general solution for Eq.~\eqref{k eq with all decay} is of the form:
\begin{equation}\label{analytical_solution_case_1}
    K(t) = \frac{\mc{N}(1-\mc{M})}{2\hat{\Gamma}G\mu t}\frac{\left[1- \left(\frac{t}{t_i}\right)^{\mc{N}\mc{M}} \right]}{1 + \left(\frac{\mc{M}-1}{\mc{M} + 1}\right)\left(\frac{t}{t_i}\right)^{\mc{N}\mc{M}}} \ ,
\end{equation}
where $\mc{N}$ is given by Eq.~\eqref{eq:N2}, $\mc{M}^2 = {1+4\hat{\Gamma}G\mu \mc{A}/\mc{N}^2}$ and $\mc{A} = F(\bar{v})\tilde{c}\bar{v}/(\alpha\xi^2)$. Note that this solution is valid for $\mc{N}\neq 0$ (or, equivalently, for $q\neq q_c$). The characteristic lengthscale of small-scale structure, however, may now achieve a linear scaling regime, with $l_k\propto t$, both for $\mc{N}>0$ (or $q<q_c$) and $\mc{N}<0$ (or $q>q_c$), but these regimes will be qualitatively different.

For $\mc{N} > 0$,  we have that 

\be
K(t) \simeq \frac{\mc{N}(1+\mc{M})}{2\hat{\Gamma}G\mu t}\,,
\ee
for $t\gg t_i$. Let us consider a value of $q=q_c-\delta_-$, with $q_c>\delta_->0$. In this case, we may write

\be 
K(t) \simeq\frac{\cc \vv \delta_-}{2\xi \hat{\Gamma}G\mu t}\left[1+\left(1+\frac{4\hat{\Gamma} G\mu F(\vv)}{\alpha \cc\vv \delta_-^2}\right)^{1/2}\right]\,.
\ee 
Assuming that the length of the loops produced is significantly larger than the gravitational backreaction scale\footnote{This is a natural expectation since gravitational backreaction smooths out cosmic strings on scales smaller than $\hat{\Gamma} G\mu L$.}, with $\alpha \gg \hat{\Gamma}G\mu$, we have that:
\begin{equation}
    K(t) \simeq \frac{\tilde{c}\bar{v}}{\hat{\Gamma}G\mu}\frac{\delta_-}{L} \longrightarrow l_k(t) \simeq \frac{\hat{\Gamma}G\mu L}{\delta_-\tilde{c}\bar{v}} \ .
\end{equation}
We then see that, in this case, the characteristic lengthscale of small-scale structure is essentially determined by the gravitational backreaction scale. The attainment of a linear scaling is therefore guaranteed by the decay of kinks due to their emission of gravitational radiation. 

When $\mathcal{N } < 0$, on the other hand, we have:

\be 
K(t) \simeq \frac{\mc{N}(1 - \mc{M})}{2\hat{\Gamma}G\mu t}\,,
\ee
for $t\gg t_i$. By writing $q = q_c + \delta_+$, with $\delta_+ >0$, one finds

\be 
K(t) \simeq \frac{ \cc \vv\delta_+}{2\xi\hat{\Gamma}G\mu t}\left[-1 +\left(1+\frac{4\hat{\Gamma}G\mu F(\vv)}{\alpha \cc \vv \delta_+^2}\right)^{1/2}\right]\,.
\ee 
We now have, up to first order in $\hat{\Gamma}G\mu/\alpha$, that:
\begin{equation}\label{k for q higher than q_c}
    K(t) \simeq \frac{F(\vv)}{\delta_+ \ell} \longrightarrow l_k(t) \approx \frac{\delta_+}{F(\vv)}\ell \ .
\end{equation}
In this case, loop production, and the consequent removal of kinks, is now sufficient to ensure that $l_k$ reaches a linear scaling regime. The characteristic lengthscale of small-scale structure is determined by the typical length of loops. This means that, typically, the linear kink density on long strings will be much smaller than in the previous case and, as we shall see, small-scale structure will have a smaller impact on the large-scale dynamics of the cosmic string network~\footnote{Note that a similar result may be found using equation Eq.~\eqref{case 1 analytical solution} for $q>q_c^*$ or $q>q_c$, depending on whether the impact of Hubble stretching is considered or not.}. The critical value of $q$ then now separates these two different scaling regimes that are achieved through two distinct physical processes, loop production and kink decay, and in which the overall impact on the evolution of the network may be, as we will see in Sec.~\ref{sec:v2s}, quite different.

For $q=q_{c}$, or $\mathcal{N}=0$, we have instead that:
\begin{equation}\label{eq. 3.35}
   t K_{\text{critical}}(t) = \left(\frac{\mc{A}}{\hat{\Gamma} G\mu}\right)^{1/2}\tanh{\left[\left(\mc{A}\hat{\Gamma}G\mu\right)^{1/2}\log\left(\frac{t}{t_i}\right)\right]} \ ,
\end{equation}
and therefore the attainment of a linear scaling regime is not possible.

\subsection{Modified loop production}\label{sec:modloop}
In this subsection, we relax the assumption that the length of loops is determined by the characteristic length of the network and consider other loop forming scenarios proposed in the literature and motivated by numerical simulations. 

\subsubsection{Loop length determined by $l_k$}
Loop production and small-scale structure are intrinsically related and, for this reason, in~\cite{allen_caldwell_q} the authors proposed that loops could be formed with a physical length that is determined by the typical interkink distance $\ell=\alpha_k l_k$, where $\alpha_k>1$ is a constant. In this scenario (we will refer to it as Case 2), every time a loop detaches $N_r=\alpha_k$ kinks are removed, unlike in the previously discussed scenario (Case 1) in which the number of kinks removed by a loop increases when $l_k$ decreases. The attainment of a linear scaling regime through loop production alone is then necessarily harder to achieve. As a matter of fact, considering just the impact of loop production, Eq.~\eqref{eq:K-master} now becomes

\be 
\frac{dK}{dt}=\cc\vv \frac{K}{L}\left[\frac{F}{\alpha_k}-q\right]+\frac{2K}{L}\frac{dL}{dt}-3HK\,,
\label{eq:eomK-case2}
\ee 
and, if one once again one assumes scaling of the network, we should have that

\be 
K(t)=K_i \left(\frac{t}{t_i}\right)^{\tilde{\mc{N}}-1}\,,
\ee 
where $\tilde{\mc{N}}=\left(F/\alpha_k-q\right)\cc\vv/\xi+3(1-\beta)$ and $K_i=K(t_i)$ is the initial kink energy density~\footnote{Note that, in this scenario, we cannot assume that $K(t_i)=0$, as this would imply that the first loops would be created with infinite length.}. From this, one may see that a linear scaling regime is only attainable if $q=\tilde{q}_c^*\equiv F/\alpha_k+q_c^*$, or equivalently, whenever the number of kinks removed $\alpha_k$ compensates exactly for the number of kinks that are created. For $q>\tilde{q}_c^*$, more kinks are removed than are created and therefore the linear kink density monotonically decreases. For $q<\tilde{q}_c^*$, not enough kinks are created and small-scale structure builds up on strings.

The picture changes however if one adds the impact of non-loop forming intercommutations to the evolution equation for $K$ (by considering, in Eq.~\eqref{eq:eomK-case2}, that the rate of kink production is given by Eq.~\eqref{intercom}). Assuming linear scaling, we again find that the general solution for $K$ is of the form of Eq.~\eqref{case 1 analytical solution}, but now with

\be
\mc{N}\to\tilde{\mc{N}}\quad\mbox{and}\quad \mc{A}\to\tilde{\mc{A}}\equiv\chi(\vv)/\xi^2
\ee
The introduction of non-loop forming intercommutations then has a significant impact in Case 2: it makes it qualitatively more very similar to Case 1. We then have scaling for $q>\tilde{q}_c^*$ as well and the scaling interkink distance is given by:

\be
l_k \simeq \delta_- \frac{\cc\vv L}{\chi(\vv)}\,,
\label{eq:lk-case2}
\ee
which, as in Case 1, is determined by the characteristic length of the cosmic string network (recall that, in case 1, the length of loops is determined by the characteristic length)\footnote{Notice that the dependence of $l_k$ on $\chi(\vv)$ is also present in Eq.~\eqref{k for q higher than q_c}, but is ``hidden" inside $F(\vv)$.}.

If one now introduces kink decay due to Hubble stretching and gravitational wave emission, the situation remains qualitatively similar to Case 1. As a matter of fact, if one assumes that the network is in linear scaling, $K$ should still be of the form in Eq.~\eqref{analytical_solution_case_1}, but now with

\be
\mc{N}\to\tilde{\mc{N}}-2\beta(1-\vv^2)\quad\mbox{and}\quad \mc{A}\to\tilde{\mc{A}}\,.
\ee
A major difference between these two loop production scenarios is that quite generally the critical value of $q$ in case 2 is larger than in case 1: we have $\tilde{q}_c=F/\alpha_k+q_c$. Only when $\alpha_k\to+\infty$ we have that $\tilde{q}_c\to q_c$. It is then harder to reach scaling without gravitational backreaction and, even when possible, the expected number density of kinks should be different. But the predictions for values of $q$ smaller than the critical value are identical in both scenarios.

Nevertheless, since we have that

\be 
\ell=\alpha_k l_k=\alpha_k \varepsilon t=\frac{\alpha_k \varepsilon}{\xi}L\,,
\ee
where we have introduced $\varepsilon\equiv L/ t$, choosing a time-dependent $\alpha$ equal to $\alpha(t)=\alpha_k \varepsilon/\xi$ would make Case 1 equivalent to Case 2. Moreover, in a linear scaling regime $\varepsilon$ and $\xi$ should remain constant and, therefore, once this regime is reached one scenario can be easily translated into the other.

\subsubsection{Two loop populations}\label{subsubsec:2pop}
One may wonder also whether these two loop formation scenarios may co-exist. As a matter of fact, Nambu-Goto numerical simulations~\cite{Lorenz:2010sm,Blanco_Pillado_2014} seem to indicate that there are two distinct loop populations: one of large loops with lengths determined by the characteristic length of the network and another of much smaller loops whose lengths may be related to small-scale structure on cosmic string and that comprise the majority of the energy loss\footnote{There is, however, a discrepancy between the number of small loops predicted in these two simulation-inferred models.}.  We will then consider a scenario in which a fraction $p$ of the loops is created with a length $\ell_2=\alpha_kl_k$, while the rest of the loops are created with a length $\ell_1=\alpha L$ (throughout this subsection, we will use the subscripts 1 and 2 to label the contribution of these two loop populations). We may then write that

\be
\frac{dn_{\rm{loops}}}{dt}=\frac{dn_{\rm{loops},1}}{dt}+\frac{dn_{\rm{loops},2}}{dt}= \frac{\cc_1\vv}{\alpha L^4}+\frac{\cc_2 \vv}{\alpha_k}\frac{K}{L^3}\,,\label{eq:loopprod-2pop}
\ee
where $\cc_1=(1-p)\cc$ and $\cc_2=p\cc$ are parameters that quantify the energy loss into loops of populations 1 and 2 respectively (notice that $\cc_1+\cc_2=\cc$). In this case, the rates of kink production and kink removal are given by

\be
\label{eq:kinkprod-case3}
\frac{dn_c}{dt}=2 \frac{d n_{{\rm loops}}}{dt} + \frac{\chi(v)}{L^4}=\left[F(\vv)-2p\right]\frac{\cc\vv}{\alpha L^4} + 2p \frac{\cc\vv}{\alpha_k}\frac{K}{L^3}\,
\ee
(where, as before, we separated the contribution of non-loop-forming intercommutations and used Eqs.~\eqref{eq:Fv} and~\eqref{eq:loopprod-2pop} in the last equality) and

\be
\label{eq:kinkremoval-case3}
\frac{dn_r}{dt}=\ell_1 q K \frac{d n_{{\rm loops},1}}{dt}+\ell_2 q k \frac{d n_{{\rm loops},2}}{dt}=\frac{q\cc\vv K}{L^3}\,.
\ee

Inserting Eqs.~(\ref{eq:kinkprod-case3}) and~(\ref{eq:kinkremoval-case3}) into Eq.~\eqref{eq:K-master} and assuming that the cosmic string network is in a linear scaling regime, we find that the linear kink density should evolve as

\bq 
\frac{dK}{dt} & = & \left[F(\vv)-2p\right]\frac{\cc\vv}{\alpha L^2}-\left(q-\frac{2p}{\alpha_k}\right)\cc\vv \frac{K}{L}+ \\ 
& + & \frac{2K}{L}\frac{dL}{dt}- HK\left[3+2(1-2\vv)\right]-\hat{\Gamma}G\mu K^2\nonumber\,.
\eq
By comparing this equation to Eq.~\eqref{k eq with all decay}, one may see that they are very similar: this scenario with two distinct loop populations reduces effectively to case 1 but with a reduced rate of kink production and a value of $q$ that is effectively reduced as well. When linear scaling is maintained as a result of the GW emission, there should be no discernible differences between this scenario and case 1. However, quite generally, when the scaling regime is maintained by loop production, it is still loop population 1 that determines the kink density, but this density is reduced. For instance, if one considers the loop production scenario measured in the simulations of~\cite{blanco2014number}, wherein population 1 is characterized by $\alpha \sim 0.34$ and $1-p=0.1$, in the radiation era, the linear kink density is reduced by a factor of $~3.2$ when compared to the case in which there is only one loop population characterized by the same $\alpha$.

Moreover, as result of the effective decrease in the relative kinkiness, the critical value of $q$ is increased as well: $q_c\to q_c+2p/\alpha_k$. So, as the fraction of loops in population 2 increases, attaining scaling only through loop production requires higher and higher $q$ (recall that achieving this type of scaling in case 2 is always harder).

\section{A velocity-dependent 2-scale model for strings with small-scale structure}\label{sec:v2s}
Up to now, we have neglected the impact of small-scale structure on the large-scale evolution of the cosmic string network. Since kinks decay by emitting gravitational radiation, however, the string network is expected to lose energy at a rate~\cite{Quashnock:1990qy, Hindmarsh:1990xi, PhysRevD.42.354}:
\begin{equation}\label{drhodt kinks}
    \left.\frac{d\rho}{dt}\right|_{\rm kinks} = - \Gamma G\mu \rho K \ .
\end{equation}
The evolution of the characteristic length of the network $L$ should then be coupled to the evolution of the characteristic length of small-scale structure $l_k$, through an additional term of the form:
\begin{equation}
  \left.\frac{dL}{dt}\right|_{\rm kinks} = \frac{1}{2}\Gamma G\mu \frac{L}{l_k} \ .
  \label{eq:dLdt kink}
\end{equation}
This modification will, in turn, result in an additional term in the evolution equation for $l_k$ (or $K$), through its dependence on $dL/dt$ (see, e.g., Eq.~\eqref{k eq with all decay}). This will result effectively in a weakening of the impact of kink decay due to gravitational radiation on the evolution of $l_k$, as $\Cb$ in Eq.~\eqref{k eq with all decay} will be reduced to $\Cb-1$. We may then describe the evolution of a cosmic string network with small-scale structure using

\bq
\frac{d\vv}{dt} & = & (1-\vv^2)\left[\frac{\kv}{L}-2H\vv\right]\,,\label{V2S-v}\\
\frac{dL}{dt} & = & (1+\vv^2)HL+\frac{1}{2}\cc\vv + \frac{1}{2}\Gamma G\mu \frac{L}{l_k}\,,\label{V2S-L}\\
\frac{dl_k}{dt} & = & 2(1-2\vv^2)Hl_k+(\Cb-1)\Gamma G\mu -\label{V2S-lk}\\
& - & (1-q)\cc\vv \frac{l_k}{L}-F(\vv) \frac{\cc\vv}{\alpha}\left(\frac{l_k}{L}\right)^2\nonumber\,.
\eq 
 So, to account for the impact of small-scale structure, we need to include (at least) one additional lengthscale in the VOS model, making it a two-scale model instead. In this section, we will use this Velocity-dependent 2-Scale model (or V2S model, for short) to study the impact of small-scale on the evolution of cosmic string networks.

Let us start by investigating whether the existence of small-scale structure may prevent the attainment of a linear scaling regime. In particular, we will investigate the existence of attractor solutions of the form
\begin{equation}\label{mod vos linear scaling}
    \bar{v} = \text{constant}\,, \quad L = \xi t \quad \text{and} \quad l_k = \varepsilon t \,,
\end{equation}
in which both lengthscales evolve linearly with physical time. Substituting these solutions into Eqs. (\ref{V2S-v})-(\ref{V2S-lk}), we find:
\bq
\bar{v}^2 & = & \frac{k(\bar{v})\left(1-\beta - \frac{\Gamma G\mu}{2\varepsilon}\right)}{\beta[k(\bar{v}) + \tilde{c}]} \,,\label{v-sca}  \\ 
\xi^2 & = &\frac{k(\bar{v})[k(\bar{v}) + \tilde{c}]}{4\beta\left(1-\beta - \frac{\Gamma G\mu}{2\varepsilon}\right)} \,,\label{L-sca}\\
\varepsilon & = & \Gamma G\mu \frac{D+(\hat{C} -1) \left[\kv +\cc\right]}{2D(1-\beta) + \left[\kv+\cc\right](1-3\beta)}\,,\label{lk-sca}
\eq
where 
\be 
\label{Dfactor}
D=\cc\left[\frac{F(\\v)}{\alpha_k}+(1-q)\right]+3\kv\,.
\ee 
In deriving Eq.~\eqref{lk-sca}, we took advantage of the fact that, as discussed in Sec.~\ref{sec:modloop}, case 1 (in which $L$ determines the length of loops) and case 2 (in which it is determined by $l_k$) are fully equivalent in a linear scaling regime, since re-writing Eq.~\eqref{V2S-lk} in terms of $\alpha_k=\alpha \xi/\varepsilon$ allows us to derive a more concise expression. Note however that, despite this, Eq.~\eqref{lk-sca} has a implicit dependence on $\vv$, $\xi$ and $\varepsilon$ through the factor $D$.

From Eqs.~\eqref{v-sca} or~\eqref{L-sca}, we may see that the cosmic string network can only attain linear scaling provided that
\be 
\varepsilon >\frac{\Gamma G\mu}{2(1-\beta)}\,.
\ee 
Using Eq.~\eqref{lk-sca}, we find that scaling then requires that

\be
\Cb>1+\frac{1-3\beta}{2(1-\beta)}\,,
\ee
which corresponds to $\Cb>0.5\,(-0.5)$ in the radiation (matter) era. This indicates that attaining a linear scaling regime is always possible in a realistic cosmological background, since quite generally we expect $\Cb\ge 1$.

\begin{figure}[t!]
    \centering
    \includegraphics[width=1\linewidth]{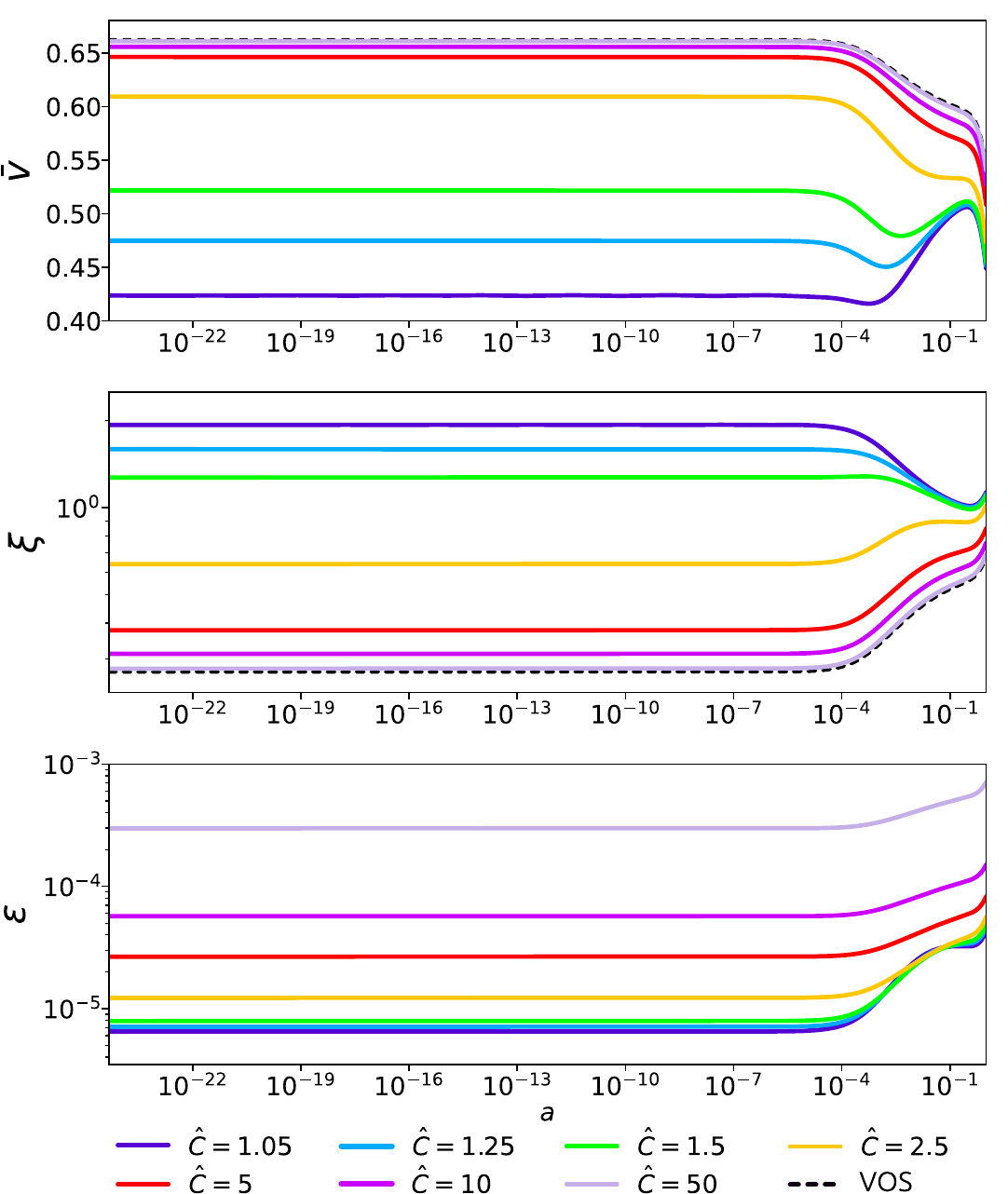}
    \caption{Evolution of a cosmic string network with small-scale structure for $q=1$ and different values of $\Cb$. The top, middle and bottom panels display, respectively, the evolution of $\vv$, $\xi$ and $\varepsilon$ in a $\Lambda$CDM cosmological background (with cosmological parameters as suggested by Planck data~\cite{Planck:2018vyg}). Here, we also took $G\mu=10^{-7}$, $\alpha=10^{-1}$, and $\cc=0.23$.}
    \label{fig:Cs}
\end{figure}

\begin{figure}[t!]
    \centering
    \includegraphics[width=1\linewidth]{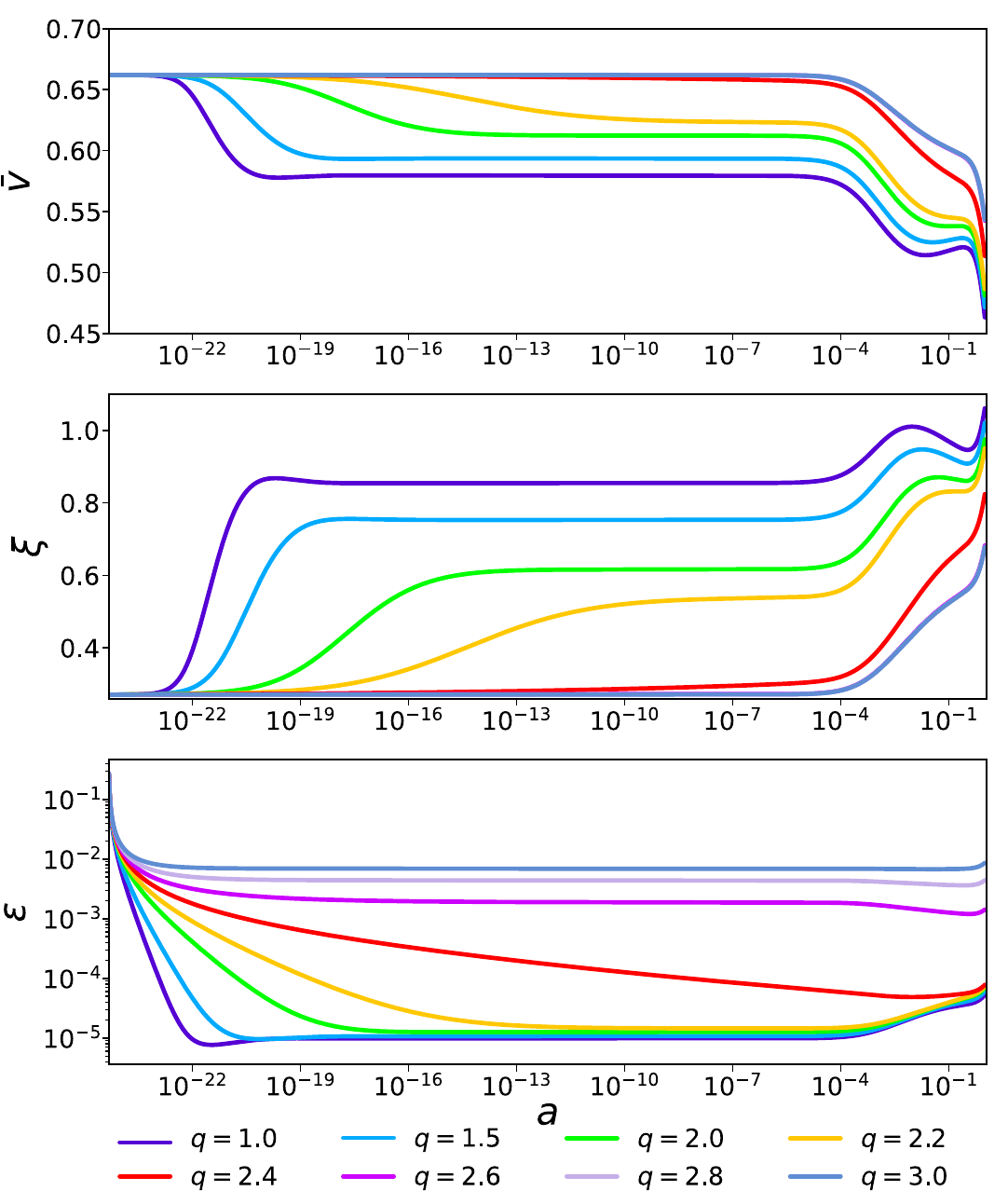}
    \caption{Evolution of a cosmic string network with small-scale structure for $\Cb=2$ and different values of $q$. The top, middle and bottom panels display, respectively, the evolution of $\vv$, $\xi$ and $\varepsilon$ in a $\Lambda$CDM cosmological background (with cosmological parameters as suggested by Planck data~\cite{Planck:2018vyg}). Here, we also took $G\mu=10^{-7}$, $\alpha=10^{-1}$, and $\cc=0.23$.}
    \label{fig:qs}
\end{figure}

As to the impact of small-scale structure, Eqs.~\eqref{v-sca} and~\eqref{L-sca} show that the effect of the energy loss caused by kink decay is to decelerate the strings and decrease their energy density (i.e., increase the characteristic length). They also indicate that the large-scale evolution of the string network will only be significantly affected by small-scale structure when $\varepsilon$ is comparable to or not much larger than the gravitational backreaction scale $\Gamma G\mu$ or, in other words, if scaling is maintained through kink decay by GW emission (and not by the loop production).

Let us then consider this limit. Since, in this case, typically $\varepsilon \sim \mathcal{O}(\Gamma G\mu)$, we have $\alpha_k \sim \mathcal{O}(\alpha\xi/\Gamma G\mu)$. Therefore, if one assumes that the length of loops is much larger than the gravitational backreaction scale, we should generally have $\alpha_k\gg1$ and, therefore, $D\simeq \cc(1-q)+3\kv$. In this case, Eqs.~\eqref{v-sca}-\eqref{lk-sca} may be written as:

\bq
v^2 & = & \frac{\kv}{\beta}\frac{\left(\Cb-\frac{3}{2}\right)(1-\beta)-\beta}{(\Cb-q)\cc+(\Cb+2)\kv}\,,\label{v-sca-gw}\\
\xi^2 & = & \frac{\kv}{4\beta}\frac{(\Cb-q)\cc+(\Cb+2)\kv}{\left(\Cb-\frac{3}{2}\right)(1-\beta)-\beta}\,,\label{L-sca-gw}\\
\varepsilon & = & \frac{\Gamma G\mu \left[(\Cb-q)\cc+(\Cb+2)\kv\right]}{\left[\kv+\cc\right](1-3\beta)+2\left[\cc(1-q)+3\kv\right](1-\beta)}\,.\label{lk-sca-gw}
\eq

Using these expressions, we may quantify the maximum impact of small-scale structure on the evolution of cosmic string networks. This corresponds to a scenario in which there is no gravitational backreaction (i.e., $\Cb=1$) and with $q=1$. In the radiation era, this corresponds to
\bq
\label{minrad}
\vv_*^r=\frac{1}{\sqrt{6}}\,,\quad\xi_*^r=\sqrt{6}k(\vv_*^r)\simeq 2.04\,,\\
\frac{\varepsilon_*^r}{\Gamma G\mu}=\frac{6k(\vv_*^r)}{5k(\vv_*^r)-\cc}\simeq 1.27\,.\nonumber
\eq
Small-scale structure may then have a very significant impact in the radiation era: the energy density may ne suppressed by a factor of up to $\sim55$, while the RMS velocity may be suppressed $\sim 1.66$. Its effect, however, is not strong enough to freeze the network and prevent the attainment of a linear scaling regime, as we previously discussed. The effect would be less significant in a matter-only universe, since the decay caused by Hubble stretching is more efficient in that case:
\bq 
\label{eq:minmat}
\vv_*^m=\frac{1}{\sqrt{4}}\,,\quad\xi_*^r=\frac{3}{2}k(\vv_*^m)\simeq 1.07\,,\\
\frac{\varepsilon_*^m}{\Gamma G\mu}=\frac{3k(\vv_*^r)}{k(\vv_*^r)-\cc}\simeq 4.43\,.\nonumber
\eq
In this case, the velocity is reduced only by a factor of $\sim 1.3$, while the energy density decreases by a factor of roughly $3$.

The impact of small-scale structure progressively decreases as $\Cb$ increases and gravitational backreaction becomes increasingly more efficient in smoothing out the kinks. This may be seen clearly in Fig.~\ref{fig:Cs}, where we display the evolution of $\vv$, $\xi$ and $\varepsilon$ for $q=1$ and different values of $\Cb$.\footnote{We do not include $\Cb=1$ as, in this limit, the integration of the V2S equations is significantly affected by numerical errors. The results obtained in the $\Cb\to1$ limit, however, clearly approach the analytical solution in Eq.~\eqref{minrad} in the radiation era.} Therein one may see that, for all values of $\Cb$, the network evolves in a linear scaling regime during the radiation era (we have verified that it indeed is well described by Eqs.~\eqref{v-sca-gw}-\eqref{lk-sca-gw} with $\beta=1/2$). On the onset of the radiation-matter transition, however, the network enters a transitional regime during which it starts evolving towards the matter-era scaling regime. The matter era, however, does not last long enough for scaling to be reestablished before dark energy becomes cosmologically relevant. From this point forward, the network starts to be quickly diluted by the fast expansion of the background, with $L,l_k\propto a$ and $v\propto 1/a$. This figure also clearly shows that, as $\Cb$ increases, the impact of small-scale structure quickly decreases and the evolution of the network becomes increasingly closer to that predicted by the VOS model.

In Fig.~\ref{fig:qs}, for visualization purposes, we have assumed that the network is already in the linear scaling regime initially. However, in a realistic setting, one would expect strings to be featureless initially and for kinks to build up dynamically as strings collide and intercommute. So, for now on, we will assume that initially the network is in the standard linear scaling regime predicted by the VOS model for strings without small-scale structure and that $l_k=L$ initially. As the purple line in Fig.~\ref{fig:qs} (representing the case $q=1$, the same value used for Fig.~\ref{fig:Cs}) shows, this has a significant impact on the evolution of the network: initially the impact of small-scale structure is negligible and the network evolves similarly to strings without small-scale structure. Only when enough structure builds up for $l_k$ to approach scaling will its impact be felt and the network will evolve towards the regime in Eqs.~\eqref{v-sca-gw}-\eqref{lk-sca-gw}. When GW emission is the mechanism that ensures scaling of $l_k$, the network will then go through two different regimes. The duration of the first (quasi-)scaling regime --- which is characterized roughly by the same $\vv$ and $\xi$ as that of bare strings --- depends on the parameters of the model. The smaller the value of $\alpha$, the shorter the duration of this initial regime will be. Moreover, for smaller values of $G\mu$, this regime will last longer, since the linear scaling regime will be characterized by a larger kink density that takes longer to attain. Its duration is, however, only weakly dependent on the value of $\Cb$.

Fig.~\ref{fig:qs} also displays the evolution of the cosmic string network for a fixed value of $\Cb$ and different values of $q$. Therein, one may see that, as $q$ increases (but remains below the critical value), the approach to scaling becomes increasingly slower and, as a result, the transient regime (in which strings are effectively kinkless) lasts longer and longer. As a matter of fact, as $q\to q_{c,r}$, the build up of small-scale structure is so slow that the radiation era may not be long enough for $l_k$ to attain scaling. We have found that, in this particular instance, this happens roughly for $2.35\lesssim q \le q_{c,r}$.  Note however that the range of values $q$ in which full scaling is not effectively attained in the radiation era depends on the other parameters of the model (namely, $G\mu$, $\alpha$, $\Cb$). For instance, if we decrease cosmic string tension to $G\mu=10^{-10}$, we find the radiation era does not last long enough to reach full scaling for $2.15\lesssim q \le q_{c,r}$. Not attaining full scaling of $l_k$, however, should not be regarded as problematic in this case: in fact, the deviations from the evolution predicted by the VOS model for strings without small-scale structure are minimal (see, for example, the line corresponding to $q=2.4$).

In this figure, one may see that, as $q$ is increases above a critical value, loop chopping becomes sufficient to ensure scaling of $l_k$ and the network then reaches a scaling regime that is characterized by a low kink density. We may also see that the critical value of $q$ is not significantly altered when one adds the energy loss caused by kink decay to the evolution of the string network and, as a result, the values of $q_{c,r}$ and $q_{c,m}$ derived in Sec.~\ref{sec:expansion} are still valid. As a matter of fact, our numerical results indicate that Eq.~\eqref{k for q higher than q_c} provides a very good description of the results if $\alpha\gg \Gamma G\mu$. We then have that, for $q>q_c$, the linear scaling regime is approximately characterized by

\bq 
\vv & \simeq & \vv_{\rm VOS}\left[1-\Delta\right]\,,\label{v-sca-q}\\
\xi & \simeq & \xi_{\rm VOS}\left[1+\Delta\right]\,,\label{xi-sca-q}\\
\varepsilon & \simeq &  \frac{q-q_c}{F(\vv_{\rm VOS})}\alpha \xi_{\rm VOS} \,, \label{lk-sca-q}
\eq
where the subscript `VOS' is used to label the values predicted by the VOS model for the corresponding quantity (obtained using Eq.~\eqref{linear scaling regime})  and, in Eqs.~\eqref{v-sca-q} and~\eqref{xi-sca-q}, we have kept terms up to first order in $\Delta\equiv \Gamma G\mu/(4\varepsilon)$. In the radiation era, we then have that

\be 
\varepsilon_r\simeq 0.126 (q-2.45)\alpha\,,\quad \Delta_r\simeq \frac{2}{q-2.45}\frac{\Gamma G\mu}{\alpha}\,,
\ee 
while in the matter era

\be 
\varepsilon_m\simeq 0.286 (q-2.67)\alpha\,,\quad \Delta_m\simeq \frac{0.9}{q-2.67}\frac{\Gamma G\mu}{\alpha}\,.
\ee 
So, the smaller the length of the loops created is, the larger the impact of small-scale structure will be. However, this effect is typically negligible in this limit, except when the length of loops approaches the gravitational backreaction scale~\footnote{How close it needs to be, however, depends on the value of $q$, as $\Delta$ also grows quite fast in the $q\to q_c$ limit.}.

\section{Discussion}\label{sec:disc}

In this paper, we have developed an effective model to describe the cosmological evolution of cosmic string networks with small-scale structure. As in~\cite{small_scale_Structure,allen_caldwell_q}, we followed a more phenomenological approach  and used two lengthscales to describe the network, the characteristic length $L$ and the typical interkink distance $l_k$. We have extended the model in~\cite{small_scale_Structure,allen_caldwell_q} by including relevant physical phenomena that have been considered in the literature --- including, for instance, the impact on non-loop-forming intercommutations~\cite{kibble_q} and the energy loss caused by kink decay~\cite{Quashnock:1990qy} --- and by treating the RMS velocity of the network $\vv$ as an additional dynamical variable. We have also incorporated the velocity-dependence into the different terms in the evolution equations for $L$ and $l_k$, as this allows for a better description of the network outside of the linear scaling regime. This is particularly important to model the evolution of the network in a realistic cosmological background (that transitions form the radiation to the matter era and later to $\Lambda$ domination) and to accurately model loop production.

This resulted in a model that, besides cosmic string tension $G\mu$ (the fundamental parameter of the model) has five additional free parameters: the GW emission efficiency $\Gamma$ (whose value has already been established through analytical and numerical studies); the loop-chopping parameter $\cc$ and the loop-size parameter $\alpha$ (both of which are fairly well known from Nambu-Goto numerical simulations); the relative kinkiness of loops $q$ and the gravitational backreaction constant $\Cb$ (which are currently unknown). All these parameters have clear physical meaning and may, in principle, be extracted from numerical simulations in the future. Note however that, at the present, numerical simulations of Nambu-Goto cosmic strings do not include the impact of gravitational backreaction on long strings\footnote{The impact of gravitational backreaction on loops has been studied in~\cite{Wachter:2024aos}.} and, as result, this is not yet possible. This obviously means that $\Cb$ and $q$ have to be treated as free parameters, but all other parameters could be affected by the inclusion of gravitational backreaction. On one hand, since the kinks created in simulations do not decay by emitting gravitational radiation, this energy loss is absent and the networks are, in principle, denser and have higher velocities than they would have in the presence of gravitational backreaction. Although Abelian-Higgs simulations also do not include gravitational backreaction, in that case there is an additional decay channel for small-scale structure on long strings, the emission of scalar and gauge radiation, and this was shown to lead to values of $L$ that are about $25\%$ higher and to values of $\vv^2$ that are about $10\%$ smaller~\cite{Hindmarsh:2017qff}. Although the nature of energy loss is quite different, this clearly shows that the decay of kinks may have a significant impact on the evolution of the network on large scales. On the other hand, this accumulation of small-scale structure on long strings in Nambu-Goto simulations should, in principle, potentiate the production of small loops (which are, as previously discussed, the vast majority of the loops measured in simulations). So, at least part of the energy that would be lost directly by the long strings may be, in current Nambu-Goto simulations, ``transferred'' to small loops and, as a result the current inferred value of $\cc$ may as well not be entirely realistic. If we again draw a parallel with Abelian-Higgs simulations (again with the caveat that the energy loss mechanism is different), the latest calibration with a VOS-type model~\cite{Correia:2021tok} indicates larger values for this parameter. Finally, loop sizes and, especially, their relative kinkiness should also be affected by gravitational backreaction. This means that, at the present, the V2S model cannot be fully calibrate, but, since simulations also have limitations and do not yet include all relevant physical processes, these semi-analytical models currently provide us with one of the few tools to study the potential impact of small-scale structure and to quantify our uncertainty as to the expected energy density of cosmic string networks (something that is crucial to make observational predictions and to accurately constrain string-forming scenarios).

The model derived in this paper shows that the impact of small-scale structure on the evolution of a cosmic string network may range from irrelevant, when loop production is enough to ensure scaling of $l_k$ (although, even in this limit, there may be some impact if $\alpha\ll 1$), to quite significant when scaling is only achieved as a result of gravitational backreaction. In this limit, the energy density of the network may be reduced, compared to the predictions of current models, by up to a factor of $55$, depending on the value of $\Cb$. Notice however that the limiting values we derived for $\vv$ and $\xi$ in the $\Cb\to1$ limit are independent of the other parameters of the model (\textit{cf.} Eqs.~\eqref{minrad} and~\eqref{eq:minmat}) and thus we may, in fact, consider this to be an absolute limiting case, in which small-scale structure has the most substantial effect.

The attainment a full scaling regime, with constant $\vv$, $\xi$ and $\varepsilon$, seems quite generally to always be possible for all values of $\Cb$ and $q$ (except for $q=q_c$). However, when loop production is insufficient to ensure scaling of $l_k$, the network will only attain this regime once enough small-scale structure has built up on the strings for $l_k$ to scale. Until this happens, the network evolves in a quasi-scaling regime, in which $\xi$ and $\vv$ are varying very slowly and assume values that are very close to those expected for bare strings, without small-scale structure. The duration of this transient regime, however, is strongly dependent on the parameters of the model --- in particular, on $\alpha$ and  $G\mu$ and, even more crucially, on $q$ --- and, in fact, for values of $q$ close to the critical value in Eq.~\eqref{eq:critical2}, it may not be possible to reach full scaling within the radiation era. Nevertheless, in this case, the deviations from the standard evolution predicted by the VOS model would always remain quite small.

\begin{figure*}[t!]
    \centering
    \includegraphics[width=0.85\linewidth]{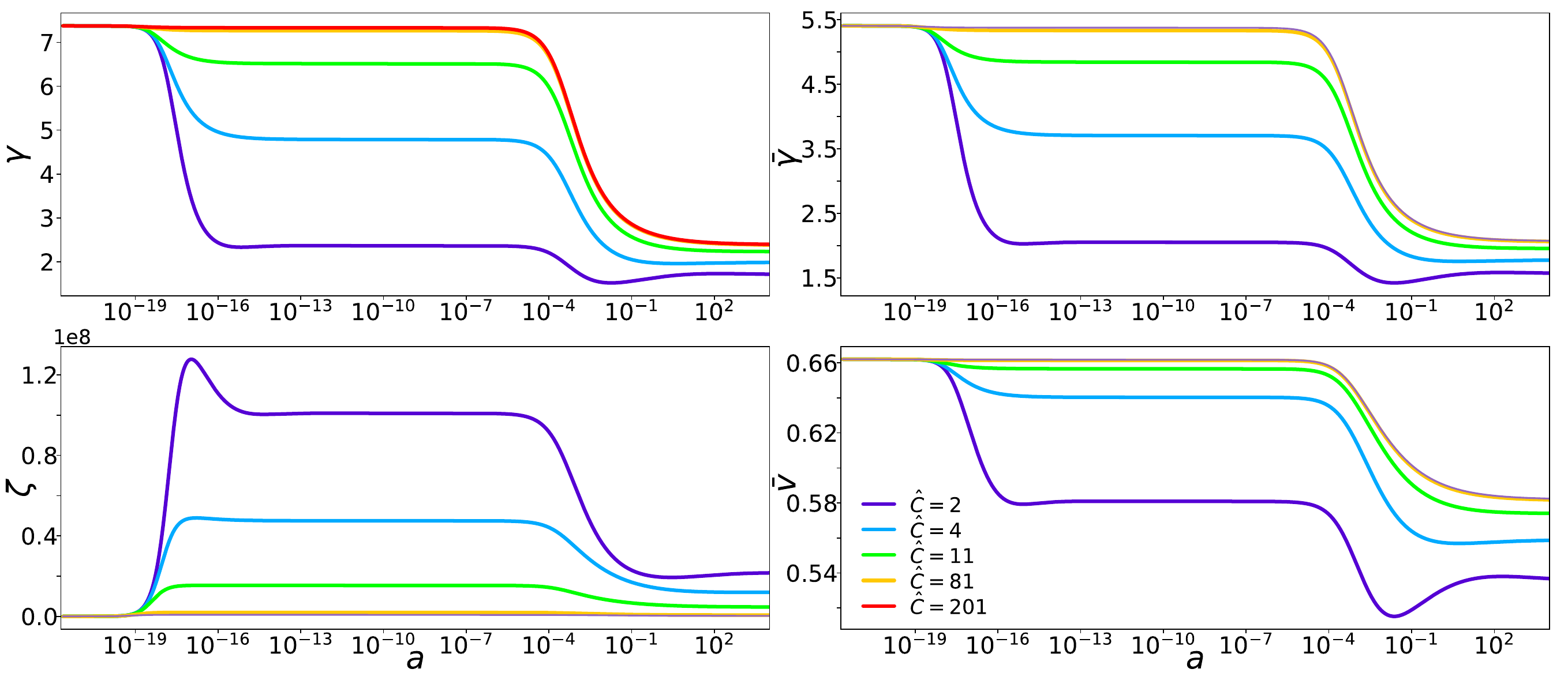}
    \caption{Evolution of a cosmic string network with small-scale structure as predicted by the V2S model derived in this paper, expressed in terms of the variables typically used in the 3-scale model in~\cite{three_scale_model,Copeland:1999gn}, $\gamma=1/(HL)$ (top left panel), $\bar{\gamma}=1/(H\lp)$ (top right panel), $\zeta=1/(Hl_k)$ (bottom left panel) and $\vv$ (bottom right panel). Here, we took $q=1.05$, $\Gamma G\mu=10^{-8}$, $\cc=0.23$ and $\alpha=0.1$ and display the results from different values of $\Cb$.}
    \label{fig:3s}
\end{figure*}

It is interesting to note that the general picture that emerges from this V2S model is quite similar to what is predicted by the 3-Scale (3S) model introduced in~\cite{three_scale_model}. The 3S model was derived using a different formalism (based on the analysis of the probability distribution of the extension of a segment of left-moving string) and includes a third lengthscale --- the persistence length $\lp$ --- in addition to the two lengthscales considered here~\footnote{Although their definition for the characteristic lengthscale of small-scale structure differs from ours, it may be roughly identified with $l_k$.}. The persistence length roughly characterizes the distance along which the direction of strings is correlated and may be roughly defined as~\cite{kibble_q,three_scale_model,Copeland:2009dk}:

\be 
\label{eq:pers}
\lp=\frac{3}{2}\frac{\varsigma}{H}\,,\quad\mbox{with} \quad \varsigma=-\langle \xx_-\cdot \xx_+\rangle\,,
\ee
where $\langle \cdots\rangle=\int\cdots \epsilon d\sigma/\int\epsilon d\sigma$ represents a weighted average over the whole network and $\xx_\pm =\xd\pm\epsilon^{-1}\xx'$ represent the left- and right-moving modes of a string in the lightcone gauge (see e.g.~\cite{Vilenkin:2000jqa}). Here, $\epsilon^2=(\xx')²/(1-\xd^2)$, $\xd$ represents the physical velocity of the string, $\xx'=d\xx/d\sigma$ its tangent vector (satisfying $\xd.\xx'=0$) and $\sigma$ is the spacelike coordinate along the string worldsheet. The quantity $\varsigma$ is related to the RMS velocity of the network $\vv=\sqrt{\langle \xd^2 \rangle}$ through $\vv^2=(1-\varsigma)/2$~\cite{kibble_q} and, as result, the evolution of the persistence length may simply be extracted from $\vv$:

\be
\label{eq:pers-v}
\lp =\frac{3}{2H}(1-2\vv^2)\,.
\ee
Using this expression, we find that the results obtained with the V2S model introduced here are in excellent qualitative agreement with the results of the 3S model. In Fig.~\ref{fig:3s}, we display the evolution of $\gamma=1/(HL)$, $\bar{\gamma}=1/(H\lp)$, $\zeta=1/(Hl_k)$ and $\vv$ for $q=1.05$, $\Gamma G\mu=10^{-8}$ and different values of $\Cb$. We chose these variables and these values fot the parameters to allow for a simpler comparison with the results of the 3S Model, as presented in~\cite{Copeland:1999gn}~\footnote{Note that some parameters in the 3S model do not have an equivalent in the V2S and that their parameter $\Cb$ corresponds to $\Cb+1$ in this model.}, and, for the same reason, we consider a universe containing only radiation and matter. Therein, one may see that the agreement between the results of the two models is indeed remarkable and that every variable displays qualitatively identical behaviour. In the case of $\gamma$, $\zeta$ and $\vv$ the results are in excellent quantitative agreement as well. The V2S model seems to predict overall smaller values for $\bar{\gamma}$, but notice that in the 3S model $\chi(\vv)$ is treated as a constant and, in the example in~\cite{Copeland:1999gn}, set to $0.1$. In the V2S model, however, we take into account its velocity-dependence and, in fact, typically for these examples we find that $\chi(v)\sim 0.23-0.24$. Since $\chi(\vv)$ enters directly in the equation for $\bar{\gamma}$ in~\cite{three_scale_model}, this may be a potential explanation for these differences. Aside from this small difference (roughly a factor of $1.3)$, these results indicate that the (deceptively) simpler evolution equation for $\vv$ includes all the relevant physical processes included in the more complex equation for $\lp$ derived in~\cite{three_scale_model}. The explanation for this lies in the form chosen for the momentum parameter, formally defined as~\cite{martins_VOS,martins_k(v)_chopping}

\be 
\vv (1-\vv^2) \frac{a k(\vv)}{L}\equiv \left\langle \epsilon^{-2}\xx''\cdot \xd\right\rangle\,.
\ee 
This parameter may also be interpreted as the mean curvature of the string along the velocity direction: curved regions of string are accelerated as a result of tension and this naturally affects the evolution of $\vv$. The curvature on a string, by measuring the rate of change of the string tangent along the string, is naturally related to its persistence length: the higher the curvature, the shorter the distance along which the strings are correlated in direction. As a matter of fact, one may show that

\be 
\vv(1-\vv^2)\frac{k(\vv)}{L}=\frac{H}{4}\left(1-\left\langle (\xx_+\cdot\xx_-)^2\right\rangle \right)\,,
\ee 
and therefore $k(\vv)$ may also be interpreted as a measure of the directional correlation of the left- and right-moving modes. The form of $k(\vv)$ in Eq.~\eqref{k(v)} --- introduced in Ref.~\cite{martins_k(v)_chopping} and constructed phenomenologically by resorting to the helicoidal string solution~\cite{Sakellariadou:1990ne} (which oscillates periodically between a straight line and a helix) --- seems then to capture appropriately the dependence of the averaged acceleration felt by the strings on $\vv$ and then provides us with an alternative way to describe the evolution of the persistence length. So, one may say that the V2S model derived here is actually a model with three lengthscales (and that the VOS model, in fact, has two different lengthscales). However, choosing $\vv$ as the third variable of the model instead of $\lp$, while incorporating the velocity-dependence into the different terms, allows not only for a simplification of the equations, but also reduces the number of free parameters necessary significantly (the 3S model has 11 free parameters).

More recently, an alternative model to describe cosmic string networks with small-scale structure was introduced in~\cite{Martins:2014kda,Vieira:2016vht}. In this model, following~\cite{Carter:1990nb}, cosmic strings with small-scale structure are effectively treated as straight elastic strings that have a higher effective mass per unit length $\mu_{\rm eff}=w\mu$ (which is now different from their effective tension $T_{\rm eff} w=\mu$). This model also has two lengthscales --- the characteristic length $L$ and a correlation length that, although defined differently, is akin to $\lp$ --- and also treats $\vv$ as a dynamical variable (but a relation between $\lp$ and $\vv$ is not established). The effective tension in this model is then given by $\mu_{\rm eff}=(\lp/L)^2$. This model also includes a mesoscopic/renormalization scale (that is not a a dynamical variable of the model), but defines the scale at which $\mu_{\rm eff}$ is defined (or, in other words, defines what is small-scale structure). This leads to additional terms in the evolution equations that depend on the multi-fractal dimension of a string segment at this scale. This quantity is modelled through a phenomenological function that yields $1$ at very small scales and $2$ on super-horizon scales as observed in Nambu-Goto numerical simulations (which crucially do not include gravitational backreaction). This model also does not fully account for the decay of wiggles as a result of the emission of gravitational radiation and, as a result, a detailed comparison is not possible. Their predictions for a constant renormalization scale~\cite{Almeida:2021ihc} are however consistent with our results (in the absence of gravitational backreaction): they find that depending on the balance between the kinks that are produced and removed, the cosmic string network may evolve towards a full scaling regime or a growing wiggliness regime. This is equivalent to the scenario we have described before Sec.~\ref{subsec:kink-decay}.

Beyond these thermodynamical models to describe small-scale structure, that are more concerned about macroscopic predictions for the network as a whole, other studies have been dedicated to the properties of cosmic strings on very small scales. Refs.~\cite{Polchinski:2006ee,Polchinski:2007rg,Dubath:2007mf} studied how small-scale structure affects loop production and their results indicate that it may happen at two distinct scales (as observed in numerical simulations): a fraction would form at a scale not much smaller than $H^{-1}$, while the vast majority would be formed around the typical small-scale structure length. We have discussed this scenario in Sec.~\ref{subsubsec:2pop} and showed that it may be straightforwardly described in our model by a simple redefinition of $F(\vv)$ and $q$. Moreover, a characterization of the distribution of kinks by sharpness, a measure of how sharp the discontinuity at the string tangent is at the kink, was performed in~\cite{Copeland:1999gn}. Their results indicate that the number of kinks with a given sharpness scales over most scales, except for very small-scales that may be irrelevant observationally.

\begin{acknowledgments}
L.S. thanks Ivan Rybak for many enlightening discussions about this subject. This work was funded by FCT - Fundação para a Ciência e Tecnologia (FCT), I.P., (\url{https://ror.org/00snfqn58}) through the Strategic Funding UID/04434/2025 and the research grant 2024.17828.PEX - \textit{Unveiling the early universe with topological defects} (\url{https://doi.org/10.54499/2024.17828.PEX}).
\end{acknowledgments}

\bibliography{apssamp}

\end{document}